\documentstyle[12pt]{article}
\makeatletter
\@addtoreset{equation}{section}
\makeatother

\newcommand{\bibi}{\bibitem}

\newcommand{\un}{1\!\!1}
\newcommand{\half}{\frac{1}{2}}

\newcommand{\lag}{\langle}
\newcommand{\rag}{\rangle}
\newcommand{\gm}{\gamma}
\newcommand{\pl}{\partial}

\newcommand{\Dm}{D_{\mu}^+}
\newcommand{\Dbm}{D_{\mu}^-}
\newcommand{\Pm}{\partial_{\mu}^+}
\newcommand{\Pbm}{\partial_{\mu}^-}

\newcommand{\ep}{\epsilon}

\newcommand{\kp}{\kappa}

\newcommand{\ph}{\phi}

\newcommand{\ch}{\chi}
\newcommand{\ps}{\psi}

\newcommand{\phat}{{\hat{p}}^2}

\newcommand{\phd}{\phi^{\dagger}}
\newcommand{\Ps}{\Psi}
\newcommand{\Ph}{\Phi}

\newcommand{\Om}{\Omega}
\newcommand{\Psb}{\overline{\Ps}}
\newcommand{\psb}{\overline{\ps}}

\newcommand{\Dsl}{D\!\!\!\!/}

\newcommand{\dsl}{\partial \!\!\!/}

\newcommand{\mn}{m_F^{(n)}}
\newcommand{\mc}{m_F^{(c)}}
\newcommand{\mdn}{m_D^{(n)}}
\newcommand{\mdc}{m_D^{(c)}}
\newcommand{\En}{E_F^{(n)}}
\newcommand{\Ec}{E_F^{(c)}}
\newcommand{\Edn}{E_D^{(n)}}
\newcommand{\Edc}{E_D^{(c)}}

\newcommand{\Pc}{\Psi^{(c)}}
\newcommand{\Pn}{\Psi^{(n)}}
\newcommand{\psn}{\psi^{(n)}}
\newcommand{\psch}{\psi^{(c)}}
\newcommand{\psbn}{\overline{\psi}^{(n)}}
\newcommand{\psbch}{\overline{\psi}^{(c)}}

\newcommand{\Pnb}{\overline{\Psi}^{(n)}}
\newcommand{\Pcb}{\overline{\Psi}^{(c)}}

\newcommand{\hmu}{\hat{\mu}}

\newcommand{\ra}{\rightarrow}

\newcommand{\be}{\begin{equation}}
\newcommand{\ee}{\end{equation}}
\newcommand{\bea}{\begin{eqnarray}}
\newcommand{\eea}{\end{eqnarray}}

\newcommand{\eq}{\ref}
\newcommand{\beq}{\begin{equation}}
\newcommand{\eeq}{\end{equation}}
\newcommand{\cc}{\cite}
\newcommand{\lb}{\label}
\newcommand{\gsim}{\stackrel{>}{\sim}}  
\newcommand{\lsim}{\stackrel{<}{\sim}} 

\newcommand{\SU}{G$_L \otimes $G$_R\;$}

\def \3{\ss}

\def\dateandnumber(#1)#2#3#4{
\vbox to 18mm{%
     \hbox to \textwidth{ \hspace*{14mm} \hsize=40mm%
            \vbox{%
                 \hbox to 40mm{\large #1 \hss}%
                 \hbox to 40mm{    \hss}%
                 \hbox to 40mm{    \hss}%
                 }%
                 \hss \hsize=80mm%
            \vbox{%
                 \hbox to 80mm{\hss \large #2}
                 \hbox to 80mm{\hss \large #3}
                 \hbox to 80mm{\hss \large #4}
                 }%
            \hspace*{14mm} }%
      \vss
    }
}
\def\titleofpreprint#1#2#3#4{{\LARGE \bf
\vbox to 43mm{%
     \vss
     \hbox to \textwidth{ \hspace*{14mm} \hsize=130mm%
            \hss \vbox{
                      \hbox to 130mm{\hss \LARGE \bf #1\hss}%
                      \hbox to 130mm{\hss \LARGE \bf #2\hss}%
                      \hbox to 130mm{\hss \LARGE \bf #3\hss}%
                      \hbox to 130mm{\hss \LARGE \bf #4\hss}%
                 }%
            \hss \hspace*{14mm} }%
      \vss
    }}
}
\def\listofauthors#1#2#3{{\large
\vbox to 22mm{%
     \vss
     \hbox to \textwidth{ \hspace*{14mm} \hsize=130mm%
            \hss \vbox{
                      \hbox to 130mm{\hss \large #1\hss}%
                      \hbox to 130mm{\hss \large #2\hss}%
                      \hbox to 130mm{\hss \large #3\hss}%
                 }%
            \hss \hspace*{14mm} }%
      \vss
    }}
}
\def\listofaddresses#1#2#3#4#5{{\small
\vbox to 18mm{%
     \vss
     \hbox to \textwidth{ \hspace*{14mm} \hsize=130mm%
            \hss \vbox{
                      \hbox to 130mm{\hss \small #1\hss}%
                      \hbox to 130mm{\hss \small #2\hss}%
                      \hbox to 130mm{\hss \small #3\hss}%
                      \hbox to 130mm{\hss \small #4\hss}%
                      \hbox to 130mm{\hss \small #5\hss}%
                 }%
            \hss \hspace*{14mm} }%
      \vss
    }}
}
\def\abstractofpreprint#1{{\normalsize
  \vbox to 110mm{%
     \vss
     \hbox to \textwidth{\hss \normalsize \bf Abstract \hss}%
     \vspace*{1cm} \normalsize
     #1
     \vss
    }}
}

\def\footnoteitem(#1)#2{
\begin{list}{#1}{\labelwidth4.0mm \leftmargin7.0mm
\labelsep2.5mm \rightmargin7.0mm \parsep0.5ex plus0.2ex minus0.1ex
\itemsep0ex plus0.2ex }
\item #2
\end{list}
}
\begin{document}
\dateandnumber(July 1992)%
{Amsterdam ITFA 92-21}%
{HLRZ J\"ulich 92-52}%
{                    }%
\titleofpreprint%
{      Fermion-Higgs model with                                }%
{      strong Wilson-Yukawa coupling                           }%
{      in two dimensions                                       }%
{                                                              }%
\listofauthors%
{Wolfgang Bock$^{1,\$}$, Asit K. De$^{2,3,\#}$,                }%
{Erich Focht$^{2,3,\&}$ and Jan Smit$^{1,*}$                   }%
{                                                              }%
\listofaddresses%
{\em $^1$Institute of Theoretical Physics, University of Amsterdam,}%
{\em \\ Valckenierstraat 65, 1018 XE Amsterdam, The Netherlands }%
{\em $^2$Institute of Theoretical Physics E, RWTH Aachen,}%
{\em Sommerfeldstr., 5100 Aachen, Germany }%
{\em $^3$HLRZ c/o KFA J\"ulich, P.O.Box 1913, 5170 J\"ulich, Germany}%
\abstractofpreprint{The fermion mass spectrum is studied in the
quenched approximation in the strong coupling vortex phase (VXS) of a globally
U(1)$_L \otimes$U(1)$_R$ symmetric scalar-fermion model
in two dimensions. In this phase fermion doublers can be completely removed
from the physical spectrum by means of a strong Wilson-Yukawa coupling.
The lowest lying fermion spectrum in this phase consists most probably
only of a massive Dirac fermion which has charge zero
with respect to the $U(1)_L$ group. We give evidence that
the fermion which is charged with respect to that subgroup is
absent in the VXS phase. When the $U(1)_L$ gauge fields are turned on,
the neutral fermion may couple chirally to the massive
vector boson state in the confinement phase. The outcome is very similar
to our findings in the strong coupling symmetric phase (PMS) of
fermion-Higgs models with Wilson-Yukawa coupling in
four dimensions, with the exception that in four dimensions the neutral
fermion does most probably decouple from the bosonic bound states.}
\begin{flushleft}
$\$$ e-mail: bock@phys.uva.nl   \\
$\#$ e-mail: hkf212@djukfa11.bitnet   \\
$\&$ e-mail: hkf247@djukfa11.bitnet   \\
$*$ e-mail: jsmit@phys.uva.nl
\end{flushleft}
\pagebreak
%
%
%
%
%
%
%
\section{Introduction}
A non-perturbative formulation of a chiral gauge theory
on the lattice has proved to be a difficult issue.
In a chiral gauge theory naively transcribed to the lattice, each fermion
is accompanied by fifteen `doubler fermions' where eight of these
couple as mirror fermions and spoil the chiral couplings.
One way to deal with this problem is to decouple the unwanted extra
species by rendering them very heavy. \\

The standard Wilson mass term, which is known to remove the doublers
in the case of vector-like theories on the lattice, obviously breaks
gauge invariance of the chiral gauge theory.
A proposal to overcome this difficulty is the so-called
Wilson-Yukawa approach \cc{Sm86,Sw84} which has recently received a lot
of attention. Instead of a standard Wilson mass term one uses
a so-called Wilson-Yukawa term which contains the Higgs fields in a way that
it is manifestly invariant under the chiral gauge transformation.
The goal of prime importance is now to try to
decouple the unwanted species doublers by means of a
strong Wilson-Yukawa coupling and to give
them a mass of the order of the cutoff. It is well known that such a decoupling
is a non-trivial and non-perturbative issue.
For recent overview articles on the Wilson-Yukawa approach see
refs.~[3-9]. \\

Recently extensive investigations
of globally \SU symmetric (with G$_{L,R}=$SU(2), U(1))
fermion-Higgs models with Wilson-Yukawa coupling in four
space-time dimensions have shown that it is rather unlikely that
this method leads to the desired non-perturbative
formulation of the standard model on the lattice.
The reasons for this may be summarized as follows:
The phase diagram contains, apart from weak
coupling symmetric (PMW) and broken (FM(W)) phases where fermion masses
behave according to perturbation theory, also strong
coupling symmetric (PMS) and broken (FM(S))
phases where these masses exhibit a non-perturbative behavior.
The Wilson-Yukawa approach fails in the weak coupling phases
because there the masses of the doubler fermions are restricted by the
triviality of Yukawa couplings and cannot be made sufficiently
heavy \cc{BoDe91a}. They remain as additional particles in the
spectrum. \\
On the other hand, in the strong coupling phases PMS and FM(S)
the doubler fermion states can be removed
completely from the particle spectrum by making them as heavy as
the cut-off. In contrast to the weak coupling region,
fermions become massive also in the PMS phase, a situation
which differs already from that in the fermion-Higgs sector of
the perturbative
standard model where the fermion mass vanishes in the symmetric phase.
Symmetry considerations show that the particle spectrum in the PMS phase
can a priori contain a fermion which is neutral with respect
to the G$_L$ group
and a fermion which is charged with respect to that group
[11-19].
The existence of the neutral fermion was confirmed by the
good agreement of analytic and numerical calculations
\cc{BoDe91,AoLe90,Le91}.
On the other hand a $1/w$ expansion ($w$ is
the Wilson-Yukawa coupling) of the charged
fermion propagator \cc{GoPeRi} and a numerical investigation of
the fermion spectrum \cc{BoDe91} gave evidence
that the charged fermion does not exist as a particle
in the spectrum, though this is not generally accepted \cc{Aoki}.
Under the assumption that the charged fermion is absent
one can show that the coupling of the neutral fermion to Higgs
and gauge fields vanishes as a power of the lattice spacing $a$ and the
neutral fermion becomes non-interacting in the scaling region
\cc{GoPe91b}. It was also argued that the renormalized
Yukawa coupling in the FM(S) phase vanishes as a power
of $a$ rather than logarithm of $a$ \cc{GoPe91b,BoDe91}. \\

In this paper we extend the investigations to a U(1)$_L \otimes$U(1)$_R$
symmetric fermion-Higgs model with Wilson-Yukawa coupling
in two dimensions. The important advantage of an investigation
in two dimensions is that
the simulations can be carried out on lattices of large linear dimension,
enabling one to achieve large correlation lengths for the scalar fields
with better control over finite
size effects. The numerical data are of much superior quality
than in the four dimensional case.\\

In two dimensions spontaneous breakdown
of a continuous symmetry cannot occur because of the
Mermin-Wagner-Coleman theorem \cc{MWC}.
In spite of that, fermions are observed to acquire a mass also in
two dimensional fermion-scalar models.
For example a $1/N$ expansion in the Gross Neveu models
with a continuous chiral symmetry
(which can be also viewed as fermion-scalar models
after the introduction of an auxiliary scalar field) shows that
the fermion mass does not vanish \cc{GrNe74}. At a first glance
this appears to be contradictory. The contradiction could be resolved
by expressing the action in terms of new fermionic fields which
can have a mass term without destroying the original chiral
symmetry \cc{Wi78}. Interestingly, the existence of massive fermions in the
strong coupling symmetric phase PMS in the four dimensional models
with Wilson-Yukawa coupling may be viewed
in a similar way \cc{Sm89,SmStAn}. The new fermionic variables
are here the above mentioned neutral and charged fermion fields
which transform vectorially under the original
chiral symmetry group and allow
therefore for the construction of invariant mass terms. \\

The phase structure of the two dimensional model in the quenched
approximation is similar to the one found before
in the four dimensional models.
The analogues of the weak coupling phases FM(W) and PMW are
respectively a weak coupling spin-wave SW(W) and
vortex VXW phase. The strong coupling
phases FM(S) and PMS get replaced by strong coupling spin-wave SW(S) and
vortex VXS phases. We find that fermions are massive
in the strong coupling phases VXS and SW(S).
The existence of the strong coupling phases is recently confirmed also by
an investigation of a two dimensional U(1) fermion-Higgs model
with dynamical naive fermions.
There are, however, some indications favoring the absence of the VXW phase
\cc{DeFo92}.
The weak coupling spin-wave
SW(W) would in this case extend down to zero Yukawa coupling.
If this is correct, the VXW phase has to be regarded
as an artefact of the quenched approximation. \\

In this paper we study the fermion spectrum in the quenched
approximation in the strong
coupling vortex phase VXS whose existence is guaranteed in the full
model with dynamical fermions. In this phase the
species doublers can be completely removed from the spectrum
and the generation of the fermion masses
is based on the same mechanism as in the PMS phase of
the four dimensional models. Also here
the fermion spectrum can a priori consist of a neutral and a charged
fermion. Similar to the four dimensional models we find strong
indications for the absence of the charged fermion.
The spectrum consists then only of the neutral fermion and the scalar
particles. We shall show in sect.~7 that
the neutral fermion may exhibit
a chiral coupling to the vector boson state in the confinement phase
after the gauge interactions are turned on. \\

The outline of the paper is as follows: In sect.~2 we introduce the model
and describe its phase structure in the quenched approximation.
Sect.~3 deals with the fermion spectrum in the strong coupling phases
SW(S) and VXS. In sect.~4 we present the results of a leading order
hopping expansion for the neutral and charged fermion propagators.
After giving a brief report on the technical details of the numerical
simulations in sect.~5, we compare in sect.~6
the numerical results for the rest energies of the charged and
neutral fermion with those obtained from the hopping expansion.
Based on the results of sect.~6 we discuss in sect.~7 different
outcomes for the physics in the VXS phase. A brief conclusion is given
in sect.~8.
%
%
%
%
%
\section{The model and its phase diagram}
The model of interest is given
by the following gauge invariant euclidean lattice action in two
dimensions:
\be
S=\sum_x {\cal L} \;,\;\; \;
            {\cal L} = {\cal L}_{\mbox{gauge}}
                     + {\cal L}_{\mbox{scalar}}
                     + {\cal L}_{\mbox{fermion}}     \lb{S}
\ee
with
\bea
{\cal L}_{\mbox{gauge}}   &=& -\frac{1}{g^2} \mbox{Re} \; U_{12 x} \;,
                              \lb{LGAUGE}   \\
{\cal L}_{\mbox{scalar}}  &=& -\kp  \sum_{\mu=1}^{2}
                               \left[ \Phi^*_x U_{x \mu}
                               \Phi_{x+\hmu}+\Phi_{x+\hmu}^*
                               U^*_{x \mu} \Phi_x
                              \right] +  \Phi_x^* \Phi_x
                               +\lambda (\Phi_x^* \Phi_x-1)^2  \;,
                               \lb{LHIGGS} \\
{\cal L}_{\mbox{fermion}} &=& \frac{1}{2} \sum_{\mu=1}^{2}
                              \Psb \gm_{\mu}
                              \left[(\Dm+\Dbm)P_L + (\Pm+\Pbm)P_R \right] \Ps
                              +y \Psb (\Phi P_R + \Phi^* P_L) \Ps
                               \nonumber \\
                          & & -\frac{w}{2} \left[ (\Psb \Phi) P_R
                               \sum_{\mu=1}^2
                               \Pm \Pbm  \Ps + \Psb P_L
                               \sum_{\mu=1}^2 \Pm \Pbm (\Phi^* \Ps)
                               \right]  \;,
                               \lb{LFERM}
\eea
where in the second line of eq.~(\eq{LFERM}) we have included the
Wilson-Yukawa coupling with strength $w$. Besides the Wilson-Yukawa
term we also included a usual Yukawa term.
The symbols $\Dm$, $\Dbm$, $\Pm$ and $\Pbm$
denote the covariant and normal lattice derivatives which are
defined by the relations
$\Dm  \Ps_x = U_{\mu x} \Ps_{x+\hmu} -\Ps_x$,
$\Dbm \Ps_x = \Ps_x - U_{\mu x-\hmu}^* \Ps_{x-\hmu}$,
$\Pm=\Dm|_{U=1}$ and $\Pbm=\Dbm|_{U=1}$.
The symbols $\gamma_{\mu}$, $\mu=1,2$ denote the two dimensional
$\gamma$ matrices and $P_{R,L}=\frac{1}{2} (\un \pm \gm_5)$ with
$\gm_5=-i \gm_1 \gm_2$ are the right and left-handed chiral projectors.
The field $U_{12x} \equiv U_{1x} U_{2 x+\hat{1}} U^*_{1 x+\hat{2}} U^*_{2 x}$
is the usual plaquette variable in the Wilson action,
$g$ is the gauge coupling, $\kappa$ is the hopping parameter
for the scalar field and $\lambda$ the quartic coupling.
If we succeed in removing the species doublers
from the particle spectrum by means of a
strong Wilson-Yukawa coupling $w$,
one would expect that the lattice
lagrangian defined by eq.~(\eq{S}) reproduces in the continuum limit the
target model given by the lagrangian
\bea
{\cal L}_{0}  &=& \frac{1}{4} F_{\mu \nu} F_{\mu \nu} +
                 D_{\mu} \phi^* D^{\mu} \phi
                 + m_0^2
                 \phi^* \phi + \lambda_0
                 (\phi^* \phi)^2 \nonumber \\
              & &+\psb (\Dsl P_L + \dsl P_R) \psi +
                 y_0 \psb (\phi P_R + \phi^* P_L ) \psi \;,  \lb{CONT}
\eea
where $F_{\mu \nu}(x)=\partial_{\mu} A_{\nu}(x)
-\partial_{\nu} A_{\mu}(x)$ and
$D_{\mu}=\partial_{\mu} -i g_0 A_{\mu}(x)$.
We will demonstrate in this paper
that this naive expectation is not correct
and that the effective lagrangian
which describes the physics in the VXS phase differs
substantially from eq.~(\eq{CONT}).
The continuum
fields $\psi(x)$, $\phi(x)$, $A_{\mu}(x)$ in eq.~(\eq{CONT})
are related to the
corresponding lattice fields in (\eq{S}) by the transformations
\be
\Psi_x= a^{1/2} \psi(x)              \;,\;\;\;
\Phi_x =\frac{1}{\sqrt{\kp}} \phi (x)     \;,\;\;\;
U_{\mu x} =\exp(-i a g_0 A_{\mu x})  \;.\lb{FIELDS}
\ee
The coupling parameters in eq.~(\eq{CONT}) can be
expressed in terms of the lattice couplings by the relations
\be
m_0^2= \frac{1-2 \lambda -4 \kp}{a^2 \kp} \;,\;\;\;
y_0=\frac{y}{a \sqrt{ \kp}}         \;,\;\;\;
g_0=\frac{g}{a}                        \;,\;\;\;
\lambda_0=\frac{\lambda}{(a \kp)^2}   \;. \lb{COUPL}
\ee
We note that in the continuum formulation
the Yukawa coupling, the gauge coupling
and the quartic coupling
carry a dimension, whereas the gauge and scalar
fields are dimensionless. Throughout this paper we will study
the model in the limit $\lambda \ra \infty$ which implies that
the scalar fields become radially frozen, i.~e. $\Phi^*_x \Phi_x =1$. \\

The lattice lagrangian in eq.~(\eq{S}) is invariant under the local gauge
transformations of the form $\Ps_{L,x} \ra \Omega_{L,x} \Ps_{L,x}$,
$\Psb_{L,x} \ra \Psb_{L,x} \Omega_{L,x}^*$,
$\Phi_x \ra \Om_{L,x} \Phi_x$, $U_{\mu x} \ra \Om_{L,x} U_{\mu x} \Om_{L,
x+\hmu}^*$, with $\Om_{L,x} \in$U(1)$_L$. The model is furthermore
also invariant under the global gauge transformations
$\Ps_{R,x} \ra \Omega_{R} \Ps_{R,x}$,
$\Psb_{R,x} \ra \Psb_{R,x} \Omega_{R}^*$ and
$\Phi_x \ra \Phi_x \Om_{R}^*$ with $\Om_{R} \in$U(1)$_R$.
At $y = 0$ the action (\eq{S}) possesses a shift symmetry
for the right-handed fermion fields,
\be
\Psi_{R,x} \ra
\Psi_{R,x} + \ep_R\;,\;\;\; \Psb_{R,x}  \ra \Psb_{R,x} +
\overline{\ep}_R\;, \lb{GP}
\ee
where $\ep_R$ and $\overline{\ep}_R$
are the constant shifts in the right-handed Weyl spinors.
This symmetry guarantees that
the mass of the fermion with the quantum numbers of the
$\Ps_R$ fermion field vanishes at $y=0$ for
all values of $\kp$ and $g$ \cc{GoPe89}. \\

Although we have included for later convenience in
(\eq{S}) also the interactions to the U(1)$_L$ gauge fields,
in our numerical work, however, we will restrict ourselves
to the global limit $g=0$ where $U_{\mu x}=1$ and the
local U(1)$_L$ gauge symmetry turns into a global one.
In this case eq.~(\eq{LHIGGS}) (with $\lambda =\infty$)
reduces to the lagrangian of the
XY model in two dimensions. We shall furthermore study the
model (\eq{S}) in the quenched approximation where the effects of
the fermion determinant are neglected. The use of the quenched approximation
for the investigation of the fermion spectrum in
strong coupling phase VXS is justified since this phase was
established also in the full model with dynamical fermions \cc{DeFo92}. \\

Next we turn to the phase structure of the model in the quenched
approximation. The XY model is known to have a
phase transition at $\kp=\kp_c \approx 0.56$ which
separates a
vortex (VX) phase ($\kp < \kp_c$) with finite scalar correlation length
from a spin-wave (SW) phase where the scalar
correlation length is infinite \cc{KoTh}. We note that the
spectrum in the VX phase
consists of two scalar particles which have the same mass.
In the quenched approximation fermions do not have a feedback on
the scalar sector and $\kp_c$ is independent of $y$ and $w$. A phase
transition may, however, occur in the fermionic sector. Such a phase
transition or `crossover' was discovered before in four
dimensional models with Wilson-Yukawa coupling
at $y+4w \approx \sqrt{2}$ \cc{BoDe89b,BaBo90}.
It separates the weak coupling phases PMW and FM(W) ($y+4w \lsim \sqrt{2}$)
from the strong coupling phases PMS and FM(S) ($y+4w \gsim \sqrt{2}$).
The existence of this `crossover' has manifested itself in a different
behavior of the fermion mass
as a function of $\kp$ in the weak and strong coupling regimes. \\

%
%
%
\begin{figure}[tb]
\vspace{12cm}
\caption{ \noindent {\em The mass of the $\Ps_R$ fermion as a function
of $y$ for several values of $\kp$ at $w=0$. The position of the phase
transition in the thermodynamic limit is given by $\kp_c \approx 0.56$.
The vertical arrow below the abscissa indicates
the position of the peak in the susceptibility on the $48 \times 48$ lattice
which lies clearly below $\kp_c$.}}
\label{td1.ps}
\end{figure}
In order to monitor the fermionic phase structure of the quenched
model in two dimensions we
have computed the mass $m_F$ of the $\Ps_R$ fermion by fitting the
$\lag \Ps_R \Psb_R \rag$ propagator to a free Wilson
fermion propagator ansatz (for more information
about the technical details see sect.~5).
In fig.~1 we have displayed this mass as a function of $\kp$ for several
values of the Yukawa coupling
$y$ and for the special case of naive fermions ($w=0$). As in four
dimensions the fermion mass shows a qualitatively different behavior at
small and large values of $y$. It decreases when approaching the
SW-VX phase transition in the
SW phase for $y < y^* \approx 1$. On the contrary
the mass is observed
to increase when the value of $\kp$ is lowered for $y>y^*$.
This increase is seen to continue
into the VX phase and fermions become massive in that phase.
A similar behavior of the physical
and doubler fermion mass as a function of $\kp$ is observed also for $w>0$.
{}From the different behavior of the fermion mass we can localize
the position of a `crossover'.
This position is within the precision of our resolution
independent of $\kp$ and is approximately given
by the relation $y+2w=1$. As in four dimensions the `crossover' sheet
splits the VX and SW phases into strong (S) and weak (W) coupling
regions, which we will denote in the following
by VXS, SW(S) ($y+2w \gsim 1$) and VXW, SW(W) ($y+2w \lsim 1$).
As we mentioned already in the introduction,
the VXW phase does not seem to be
present in the full model with dynamical fermions and could be
an artefact of the quenched
approximation \cc{DeFo92}. However, this does not affect our
investigations in the VXS phase. \\

Our numerical results on the $48^2$ lattice
show that the fermion mass stays non-zero everywhere in the SW phase
and also in the VXS phase though the
chiral symmetry cannot be broken according to the Mermin-Wagner-Coleman
theorem. Studies on different lattices show that the finite size
dependence of the fermion mass is extremely small and it appears very
unlikely that the fermion mass could vanish in the infinite
volume limit. The existence of a massive neutral fermion in the
strong coupling phases SW(S) and VXS is indeed strongly
supported by the good agreement between analytic calculations which are
based on strong coupling expansions and
the results of the numerical simulation.
We will report more on these results
in the following sections of this paper.\\

The fact that fermions may be massive within the SW phase
although the chiral symmetry is unbroken in that phase
was to our knowledge first explained in ref.~\cc{Wi78}. The
basic idea is as follows: The
original action in terms of the $\Psi$ fields does not provide a correct
description of the physics in the SW phase. It can, however, be rewritten
in terms of new fermionic field variables, which
transform vectorially under the original symmetry transformations
thereby allowing for the construction of a chirally invariant
mass term. This new form of the action may give a more appropriate
description of the physics in this phase (in the sense of a weak coupling
expansion)
if the fermions are indeed observed to be massive.
%
%
%
\section{The fermion spectrum in the strong coupling phases}
Since the U(1)$_L \otimes$U(1)$_R$ symmetry is unbroken everywhere,
the states excited by
the fields $\Psi_L$ and $\Psi_R$ need not be the same since the
corresponding fields carry different quantum numbers under the unbroken
symmetry group.
We will refer to the $\Ps_R=\Pn_R$ field as the neutral ($n$)
fermion field since it has charge zero under the U(1)$_L$ group ($q_L=0$).
The $\Ps_L=\Pc_L$ field will be called
charged ($c$) fermion field since it has charge one under
the local U(1)$_L$ gauge group ($q_L=1$). When discussing the phase diagram
in the previous section
we have already mentioned that the numerical results give strong
evidence that the
$\Ps_R$ fermion becomes massive both in the SW(S) and VXS phases.
This will be substantiated later by the good agreement of the
numerical results with a hopping expansion.
In order to describe this situation the $\Pn_R$ field may be
regarded as the right-handed component of a massive Dirac fermion field $\Pn$.
A possible choice for $\Pn_L$ is the composite field $\Phi^{*} \Ps_L$
which transforms in the same manner as $\Pn_R$.
The neutral Dirac fermion field may then be written as
\be
\Pn = ( \Phi^{*}P_L + P_R ) \Ps\;\;,\;\;\;\;\;
\Pnb =\Psb (\Phi P_R + P_L) \;.
\lb{NEU}
\ee
Along the same lines we may also introduce a charged Dirac fermion
field
\be
\Pc = (P_L + \Phi P_R) \Ps \;\;,\;\;\;\;\;
\Pcb=\Psb (P_R + \Phi^{*} P_L) \;.
\lb{CHA}
\ee
The fields $\Pc$ and $\Pn$ transform vectorially under U(1)$_L$ and
U(1)$_R$ respectively.\\

On a finite lattice the long range fluctuations in the SW phase cause
a non-vanishing value of the magnetization $M$ which
may be defined by the relation
\be
M=\lag \frac{1}{V} \sum_x \Phi_x \rag|_{\mbox{rot}}  \;, \lb{MAG}
\ee
where $V$ is the lattice volume. The index ``rot'' means
that each configuration is rotated such that
$\frac{1}{V} \sum_x \Phi_x$ points into a given direction.
This rotation is necessary since on a finite lattice
the magnetization $M$ is drifting through the group space and when averaging
over many configurations one would get zero. A measure for the magnetization
in the SW phase may then be defined by $v=|M|$.
Since there is no spontaneous symmetry breaking in two dimensional systems,
this quantity has to vanish in the infinite volume limit $V \ra \infty$.
Some typical values for $v$ on a $48 \times 48$ lattice are given by
$v(\kp)=0.3545(44)$, $0.5565(22)$, $0.6734(12)$ for
$\kp=0.48$, $0.52$, $0.60$. Even on large lattices (e.~g. $400^2$)
$v$ is clearly non-zero in the SW phase and
increases when raising the values of $\kp$. Therefore, on a finite lattice the
situation in the SW phase is very similar to
the broken phase in the four dimensional model
where the U(1)$_L \otimes$U(1)$_R$ symmetry
is broken to the diagonal subgroup U(1)$_{L=R}$.
As a consequence the fields $\Pn$ and $\Pc$ appear to
behave almost as equivalent interpolating fields.
Indeed, the numerically found rest energies obtained from the
neutral and charged fermion propagators coincide
in that phase within the statistical errors. In
the infinite volume limit, however, the two
rest energies are expected in general to be different for $w>0$.
%
%
%
\section{Hopping \hspace{0.07cm} expansion \hspace{0.07cm} for the
\hspace{0.07cm} neutral \hspace{0.07cm} and \hspace{0.07cm}
char\-ged fermion propagators}
An appropriate method to try to find analytic expressions for the neutral
and charged fermion propagators in the strong coupling     phases is the
hopping expansion. The hopping expansion deals only with   the fermionic
integration in the path integral, the bosonic      integration has to be
performed e.~g. by a $1/d$  expansion \cc{GoPed} or by numerical simulations.
When starting from    the     lagrangian   (\eq{LFERM}) with the single-site
Yukawa-coupling and expanding the Boltzmann factor in the path integral
in powers of the hopping parameter
$\alpha = 1/(y+2w)$ one comes upon cancellations of the type
$\Phi^* \Phi \ra 1$, emerging from the single-site
terms and the one-link terms.
The hopping expansion becomes more transparent after removing
the $\Phi$ fields from the
single-site Yukawa term
by means of a unitary transformation
to the one link terms.
Two transformations of this type are given by the inverses of
eqs.~(\eq{NEU}) and (\eq{CHA}) which express the
neutral and charged fermion fields in terms of the original $\Ps$ fields.
The associated jacobian for these transformations
is in both cases equal to one since the scalar field is radially
frozen. Replacing $\Ps$ and $\Psb$ in (\eq{LFERM}) by $\Pn$
and $\Pnb$ gives
\bea
{\cal L}_F &=& \half \sum_{\mu=1}^2 \left[
(\Pnb_L \Phi^*) \gm_{\mu}(\Dm+\Dbm) (\Phi \Pn_L) +
\Pnb_R \gm_{\mu}(\Pm+\Pbm) \Pn_R \right] \nonumber \\
& & +y \Pnb \Pn - \frac{w}{2} \Pnb \sum_{\mu=1}^2 \Pm \Pbm \Pn \;.\lb{LN}
\eea
This substitution transforms the Yukawa term into a bare mass and the
Wilson--Yukawa term into a free Wilson term.
Expressing the $\Ps$ and $\Psb$ fields in terms of
the $\Pc$ and $\Pcb$ fields leads to
\bea
{\cal L}_F &=& \half \sum_{\mu=1}^2 \left[
\Pcb_L \gm_{\mu}(\Dm+\Dbm) \Pc_L +
(\Pcb_R \Phi) \gm_{\mu}(\Pm+\Pbm) (\Phi^* \Pc_R )\right] \nonumber \\
& & +y \Pcb \Pc - \frac{w}{2} (\Pcb \Phi)
\sum_{\mu=1}^2 \Pm \Pbm (\Phi^* \Pc) \;.\lb{LC}
\eea
The lagrangian (\eq{LN}) has a shift symmetry (\eq{GP}) in terms
of the neutral field because $\Pn_R= \Ps_R$.
Such a symmetry, however, is absent for the action
(\eq{LC}) in terms of the $\Pc$ fields.
This different behavior under the shift symmetry
holds also if the local U(1)$_L$ gauge interactions are turned off.
It makes therefore sense
to distinguish between the charged and the neutral fermion also in
this globally symmetric U(1)$_L \otimes$U(1)$_R$ theory. \\

Using the lagrangians (\eq{LN}) and (\eq{LC}) one finds
to lowest order in $\alpha$ the following expressions
for the charged and neutral fermion propagators in
momentum space,
\bea
S^{(n)}(k)&\equiv&
\left \lag \frac{1}{V} \sum_{x,y} \Pn_x \Pnb_y e^{i k(x-y)}
\right \rag \nonumber\\
& \approx &(z^{-1}P_L +P_R)
[ (y+2w-w\sum_{\mu=1}^2 \cos k_{\mu})z^{-1}
+i \sum_{\mu=1}^2 \gm_{\mu} \sin k_{\mu} ]^{-1}
(z^{-1} P_R +P_L)  \;, \nonumber \\
& & \lb{propn}
\eea
and
\bea
S^{(c)}(k)&\equiv&
\left\lag \frac{1}{V} \sum_{x,y} \Pc_x \Pcb_y e^{i k(x-y)}
\right\rag\nonumber\\
&\approx&(P_L+ z^{-1}P_R)
[ (y+2w)z^{-1} - wz\sum_{\mu=1}^2 \cos k_{\mu}
+i \sum_{\mu=1}^2 \gm_{\mu} \sin k_{\mu} ]^{-1}
(P_R + z^{-1} P_L) \;. \nonumber \\
& &  \lb{propc}
\eea
Here
\be
z^2  = \lag \mbox{Re} (\Ph_x^{*} U_{\mu x} \Ph_{x+\hmu}) \rag   \lb{LINK}
\ee
is the scalar field link expectation value. This quantity has
a non-vanishing value in the VX as well as SW phase. \\

{}From the expressions (\ref{propn}) and (\ref{propc}) one can
read off expressions for the fermion  masses.
For the masses of the neutral fermion and its species doublers we obtain
\be
\mn  \approx  y z^{-1}\;,\;\;\;\;
\mdn \approx \mn + 2 w l z^{-1} \;,\;\;\; l=1,2 \;,
\lb{MN}
\ee
where $l$ is the number of momentum components equal to $\pi$ in the
two dimensional Brillouin zone. From eq.~(\eq{MN}) we can read off
an effective Wilson $r$-parameter:
\be
r^{(n)} \approx  w z^{-1} \lb{RN}  \;.
\ee
As we shall see later, these expressions are in
good agreement with the numerical simulations in
the VXS phase. This was  also  found to be the case in the four
dimensional models \cc{BoDe91,AoLe90,Le91}.  \\

In agreement  with the  shift symmetry \cc{GoPe89} the mass $\mn$ of
the physical fermion is seen to vanish in the limit  $y \ra 0$.
The doubler fermions,
however, are non-zero within the strong coupling
phases, since $2wlz^{-1} \neq 0$ everywhere in this region. This means
that in the continuum limit
the species doublers for the neutral fermion decouple from
the particle spectrum within this phase.  \\

Similar formulas can be obtained from eq.~(\eq{propc})
for the masses of the charged fermion (assuming it exists for the moment)
and its species doublers
\be
\mc  \approx (y+4w) z^{-1} -4wz \; ,\;\;\;
\mdc \approx \mc + 2 w z l \; ,\;\; l=1,2 \;.
\lb{MC}
\ee
The effective Wilson parameter is now given by
\be
r^{(c)} \approx  w z \lb{RC}  \;.
\lb{WC}
\ee
The discussion in the following section will show that the formulas
(\eq{MC}) and (\eq{WC}) for the charged fermion are
in disagreement with the numerical results.
A calculation of higher order terms in four dimensional models
showed that they appear to be small for the
neutral propagator $S^{(n)}$, but not for the charged propagator
$S^{(c)}$ \cc{Ao91,Le91}. On the basis of our numerical results
we will give in sect.~7 an argument
why the hopping expansion to lowest orders leads to a wrong
result for charged fermion propagator. In sect.~6 we will
compare the analytic
formulas that have been derived in this section with the
results of the numerical simulation.
\section{Details of the numerical simulation}
The neutral and charged fermion propagators were
determined by inverting the corresponding fermion matrices on a set
of uncorrelated scalar field configurations which were generated
by means of the reflection cluster algorithm \cc{Wo} for the XY model.
We have computed the fermion propagators in coordinate space.
As an example we give the expression for the
RR component of neutral fermion propagator
\be
S^{(n)}_{RR} (t)=\left\lag \frac{1}{L}
\sum_{x_1} {\Pn_{x_1,x_2}}_R {\Pnb_{y_1,y_2}}_R
 e^{ip_1 (x_1-y_1)}\right\rag \;,\;\;\; t=|x_2-y_2| \;,  \lb{ANS}
\ee
where $t=1,\ldots,T$. The symbols $L$ and $T$
denote here and in the following the spatial and time extent of a
rectangular lattice. The physical fermion propagator is obtained for
$p_1=0$ and the propagator of the lowest lying doubler
fermion for $p_1=\pi$.
The fermion and the doubler fermion propagators
were computed for all four $L$-$R$
combinations. We have used for the fermion fields periodic boundary conditions
in the spatial direction and anti-periodic boundary conditions
in the time direction. The scalar fields had periodic boundary
conditions in all directions. \\

For the neutral fermion propagator we have inverted the fermion matrix on
typically 1000-2000 scalar field configurations. A problem which we were
confronted with in the four dimensional models was the large number
of matrix inversions which was required to obtain a stable
signal for the charged fermion propagator in the PMS phase.
In the two dimensional models it is possible to enlarge the
statistics
in the VXS phase on relatively large lattices (e.~g. $32^2$)
by an order of magnitude. For the determination of the charged fermion
propagator we have inverted the fermion matrix on typically
1000-5000 and deeper in the VXS phase on 20000 scalar field
configurations. \\

Most of the results were obtained on a $32 \times 32$ lattice.
In order to estimate the finite size effects we varied $L$
at a particular point in the VXS
phase from $16$ to $64$ while keeping $T$ fixed at $64$. \\

We have fitted the neutral and charged propagators at zero spatial momentum
to the free Wilson fermion ansatz
\bea
& & S^{(n,c)}(t) \ra
\frac{Z}{2\sqrt{1+2 r_l m_l+ m_l^2}} \times  \nonumber \\
& & \left[
\frac{\exp(-E_l t) +\zeta  \exp(-E_l (T -t))}
{1+ \exp(-E_{l} T)}
- \zeta  (-1)^t \; \frac{ \exp (-E_l^{\prime} t)+ \zeta
\exp (-E_l^{\prime} (T -t)) }
{ 1+\exp (-E_l^{\prime} T) }
\right ]\;, \nonumber \\
& &  \label{ENY}
\eea
where $\zeta=1(-1)$ for the RR and LL (RL and LR) components.
This ansatz holds only for $r_l<1$, for $r_l>1$ the oscillating
factor $(-1)^t$ has to be omitted.

{}From this fit we obtain the numerical values for the rest energies
$E_l$ and $E_l^{\prime}$ of the fermion and its `time doubler' and for the
wave-function renormalization constant $Z$. The subscript
$l$ is $0$ for the physical fermion propagator ($p_1=0$ in eq.~(\eq{ANS}))
and $1$ for the doubler fermion propagator ($p_1=\pi$).
The masses $m_l$ and the
Wilson parameters $r_l$ for which we obtained expressions in the previous
section are related to the rest energies $E_l$ and $E_l^{\prime}$
by the lattice dispersion relations
\be
e^{E_l} =
\frac{\sqrt{1+2 r_l m_l+m_l^2}+r_l+m_l}{1 + r_l}\;,\;\;
e^{E_l^{\prime}}=
\frac{\sqrt{1+2 r_l m_l+ m_l^2}+r_l+m_l}{1-r_l}\; .
\label{DIS0}
\ee
The effective Wilson parameter
$r_l$ could be in principle a function of $l=1,2$.
The numerical results, however, show that $r_l$ is independent of $l$
(therefore we will subsequently use the notation $r=r_l$), which
supports the interpretation of the numerical results in terms of the free
fermion formula.

We find that the rest energies obtained from the
four chiral components $S^{(n)}_{LL} (t)$, $S^{(n)}_{RR} (t)$,
$S^{(n)}_{RL} (t)$ and $S^{(n)}_{LR} (t)$ for $p_1=0,\pi$ agreed
always within the statistical errors. The same holds also for the rest
energies determined from the four chiral components of the charged fermion
propagator. In the following we will use the notation $\En$ and $\Ec$ for the
rest
energies of the neutral and charged
physical fermion and similarly $\Edn$ and $\Edc$ for the rest energies
of the lowest lying doubler fermions.

%
%
%
\begin{figure}[tb]
\vspace{12cm}
\caption{ \noindent {\em The physical charged fermion propagator
$S^{(c)}_{RL} (t)$ ($p_1=0$) as a function of the time coordinate $t$ for
$(\kp,y,w)=(0.45,0.3,2.0)$. The lattice is size is $32 \times 64$.
The solid line was obtained by fitting $S^{(c)}_{RL} (t)$ to the free
Wilson fermion ansatz in eq.~({\protect \eq{ENY}}).}}
\label{td2.ps}
\end{figure}
In fig.~2 we have displayed as an example the charged fermion propagator
$S^{(c)}_{RL} (t)$ for $p_1=0$ as a function of $t$.
In this example we have chosen $L=32$ and $T=64$. The solid curve was
obtained by fitting the numerical data to the free Wilson fermion
ansatz (\eq{ENY}).
The high quality of the numerical results for the propagators allowed for
an accurate determination of the rest
energies $\En$, $\Edn$, $\Ec$ and $\Edc$. \\

For the considerations in sect.~7 we have to know also the
numerical values for the rest energy $E_{\Phi}$ of the scalar
particles in the VX phase. This rest energy was determined numerically
by fitting the scalar propagator in momentum space defined by
\be
G_{\Phi}(p) = \left\lag \frac{1}{2V} \sum_{x,y}
\Phi_x^* \Phi_y e^{ i p(x-y)} \right\rag
\ee
to a free scalar propagator ansatz
\be
G_{\Phi}(p) \ra \frac{ Z_{\Phi} }{ \phat + m_{\Phi}^2 } \;, \lb{FB}
\ee
where $\phat=2 \sum_{\mu=1}^2 (1-\cos p_{\mu})$ is a
lattice equivalent of the
momentum squared in the continuum. The rest energy $E_{\Phi}$
of the scalar particles in the VX phase is obtained from the lattice
dispersion relation at zero momentum, $\cosh E_{\Phi}=1+m_{\Phi}^2/2$.
%
%
%
%
\section{Comparison of the numerical results with the
hopping expansion}
In this section we are going to compare the numerically found
values for the rest energies $\En$ and $\Ec$
of the neutral and the charged fermion and the corresponding
rest energies $\Edn$ and $\Edc$ for the lowest lying doubler fermions
with the analytic expressions which result from the formula
(\eq{DIS0}) after inserting eqs.~(\eq{MN}), (\eq{RN}),
(\eq{MC}) and (\eq{RC}) from the lowest order hopping expansion.
For $z^2$ we use the numerical value measured on the same
lattice which we are using for the determination of the
fermion propagators. In fig.~3 we have displayed the numerical values
for $\En$, $\Edn$, $\Ec$ and $\Edc$ as a function of $y$
for $\kp=0.4$ and $w=2.0$.
The coupling parameter values we have
chosen here lie well inside the VXS phase. The
results from the hopping expansion are represented by the
curves. The dashed, solid, dash-dotted and dotted lines
correspond respectively to the
rest energies $\En$, $\Edn$, $\Ec$ and $\Edc$.
The figure shows that the agreement between the numerical result and
the analytic prediction is quite
impressive for the rest energies $\En$ and $\Edn$
while the curves for $\Ec$ and $\Edc$ exhibit a strong
deviation from the numerical results. In the case of $\Ec$ the deviation
is larger than a factor two. The figure shows furthermore that
$\En$ appears to vanish in the limit $y \ra 0$, in agreement with
the shift symmetry mentioned before,
whereas $\Edn$ stays non-zero for all different values
of $y$ which implies the decoupling of the species doublers
of the neutral fermion in the continuum limit.
Also the numerical values for $\Edc$
are larger than one for all values of $y$
with no indication of dropping in the limit $y \ra 0$.
Thereby also the species doublers of the charged fermion can be
completely removed from the physical spectrum. Provided
it exists at all as a particle in the spectrum, the charged fermion is
massive in the VXS phase. \\

%
%
%
\begin{figure}[tb]
\vspace{12cm}
\caption{ \noindent {\em The rest energies $\En$, $\Edn$, $\Ec$ and
$\Edc$ as a function of $y$ for $\kp=0.4$ and $w=2.0$.
The dashed, solid, dash-dotted and dotted
curves represent respectively the
analytic results for $\En$, $\Edn$, $\Ec$ and $\Edc$ obtained
from the hopping expansion.}}
\label{td3.ps}
\end{figure}

In fig.~4 the rest energies
$\En$ and $\Ec$ are plotted as a function of $\kp$ where the coupling
parameters $y$ and $w$ were fixed to $0.4$ and $2.0$.
The analytic results for
$\En$ and $\Ec$ are represented also in this figure
by the dashed and dash-dotted lines. We find again that
analytic expressions from the
hopping expansion provide a good description of the rest
energy of the neutral, but not of the charged fermion.
The other details in this figure will be explained in the next
section where we will develop,
on the basis of the results of this section,
two different scenarios for the physics in the strong coupling region.
%
%
%
\begin{figure}[tb]
\vspace{12cm}
\caption{ \noindent {\em The rest energies $\En$ and $\Ec$
plotted against $\kp$ for $y=0.4$ and $w=2.0$.
The dashed and dash-dotted curves represent the
analytic results from the hopping expansion for
$\En$ and $\Ec$. The solid line gives the result for the sum
$\En+E_{\Phi}$. The error bars for $\En+E_{\Phi}$
are much smaller than the symbol sizes
for $\Ec$.}}
\label{td4.ps}
\end{figure}
%
%
%
%
%
%
\section{Scenarios for an effective field theory in the strong coupling regime}
The results of the previous section showed that
the analytic results deduced from the lagrangian
(\eq{LN}) are in good agreement with the numerical results for
the rest energies $\En$ and $\Edn$, whereas
the lagrangian (\eq{LC}) led to expressions which were
in a strong disagreement with the numerical data. This suggests that
the physics in the strong coupling region is well described
by the lagrangian (\eq{LN}) in terms of the neutral fermion fields.
The charged fermion fields $\Pc=\Ph\Pn$ and
$\Pcb=\Pnb \Phi^*$ can then be regarded as composite fields and
the charged fermion, provided it exists at all in the particle
spectrum, can be considered as a bound state of the scalar particle
and the neutral fermion. The question arises now whether the
interactions in eq.~(\eq{LN}) can produce such a $\Phi$-$\Pn$
bound state. In the four dimensional model the scalar fields $\Phi$
carry a dimension of a mass and as a consequence of this the
four-point coupling
\be
 \half \sum_{\mu=1}^d
(\Pnb_L \Phi^*) \gm_{\mu}(\Dm+\Dbm) (\Phi \Pn_L) \lb{FOURP}
\ee
with $d=4$ vanishes in the classical continuum
limit like $a^2$ which makes the
formation of a $\Phi$-$\Pn$ bound state already very unlikely.
Indeed a $1/w$ expansion \cc{GoPeRi} and the numerical simulations
\cc{BoDe91} gave strong evidence for the absence of the charged fermion
in the particle spectrum of the PMS phase.
In the case of the two dimensional model the scalar fields are
dimensionless and for $d=2$ the coupling (\eq{FOURP})
does not vanish as a power of the lattice spacing $a$.
Therefore the formation of a $\Phi$-$\Pn$ bound state
appears at the first glance to be more favored
than in the four dimensional model.
If the four-point interaction (\eq{LN}) is strong enough to produce
a $\Phi$-$\Pn$ bound state we expect to find the following relation
among the rest energies of the neutral and the charged fermion and of
the scalar particles in the VXS phase
\be
\Ec=\En+E_{\Phi}+\ep_B   \;,\;\;\; \ep_B<0   \;,
\ee
where the quantity $\ep_B$ denotes the
binding energy of the $\Phi$-$\Pn$ bound state. This relation implies
that the charged fermion could only exist as a particle at a point
in the coupling parameter space where the rest energies
$\Ec$, $\En$ and $\ep_B$ scale simultaneously to zero.
This can happen only at the point $\kp=\kp_c$,
$y=0$, since only there $\En$ and $E_{\Phi}$
can vanish simultaneously. \\

\noindent {\bf Scenario A}: Let us assume now for the moment that
$\ep_B$ scales to zero like $\En$ and $E_{\Phi}$ in the limit
$\kp \nearrow \kp_c$, $y \ra 0$ and the charged fermion
exists together with the neutral fermion as a Dirac fermion
in the particle spectrum. The coupling of the $\Pc_L$ field to the
gauge fields is necessarily vectorial because
the charged fermion is massive ($m^{(c)} >0$).
The model we end up with in the strong coupling VX phase
is clearly different from the original target model
given in eq.~(\eq{CONT}), although we succeeded in removing the
species doublers from the spectrum.
One possible form of an effective action in the VXS phase
is given by the expression
\bea
{\cal L}_F^{eff} &=& \psbn \dsl
\psn + \psbch \Dsl \psch
+ m^{(n)}\psbn\psn + m^{(c)}\psbch\psch \nonumber\\
&&\mbox{} + y_{R} [\psbn_R\phi^* \psch_L + \psbch_L\ph \psn_R]\;,
\label{mirror}
\eea
where we used the concise continuum notation of eq.~(\eq{CONT}) and
all fields and coupling parameters are considered to be effective.
The symbol $y_R$ denotes the renormalized Yukawa coupling.
The failure of the Wilson-Yukawa approach in giving a chiral gauge theory
is related to the fact that the original fermion fields
$\Ps_R$ and $\Ps_L$ combine
with composite `mirror' fermion fields $\ch_R \equiv \ph \ps_R$ and
$\ch_L\equiv\phd\ps_L$ and form the
two massive Dirac fields $\Pn$ and $\Pc$
which would couple vectorially to ``external'' gauge fields.
The model (\eq{mirror}) is a special case of the mirror fermion model
\cc{Mo87}, when transcribed to the case
of two dimensions. It has, however, less flexibility in tuning
coupling parameters. \\
Eq.~(\eq{mirror}) is, however, not the
only possible form of an effective action in the strong coupling phase.
For example it allowed by the symmetries to add a term,
which couples the neutral fermion
chirally to the massive vector bosons in the confinement phase. We will
come to this in the last part of this section.  \\

%
%
%
\begin{figure}[tb]
\vspace{12cm}
\caption{ \noindent {\em The rest energies $\Ec$
and $\Edc$ as a function of
$y$ for $\kp=0.4$ and $w=2.0$. The
rest energies are compared with the
sums $\En+E_{\Phi}$ and $\Edn+E_{\Phi}$ which are represented
by the lines. The error bars of
$\En+E_{\Phi}$ and $\Edn+E_{\Phi}$ are much smaller than the symbol sizes
for $\Ec$.}}
\label{td5.ps}
\end{figure}

\noindent{\bf Scenario B}:
In order to figure out whether the above requirements for
the binding energy are fulfilled we have determined
$\ep_B$ numerically in a wide range
of the bare parameters in the VXS phase. The
details about the numerical determination of
the rest energy $E_{\Phi}$ were given in sect.~5.
In the figs.~4 and 5 we compare the rest energy $\Ec$
with the sum $\En+E_{\Phi}$ which in these
graphs is represented by the solid lines.
For $\En$ we have used the results
from the hopping expansion which are
in perfect agreement with the actual
data, as we reported in sect.~6.
The error bars for the sum $\En +E_{\Phi}$ are always
much smaller than the symbol sizes for $\Ec$.
Both plots indicate that
$\ep_B=0$ for all values of $\kp$ and $y$ in the VXS phase. This
suggests that the formation of a bound state is very unlikely.
Fig.~5 shows that also for the lowest lying doubler fermion
the numerical results for the
rest energy $\Edc$ are nicely
represented by the sum $\Edn+E_{\Phi}$ (upper solid line).
In order to get an impression about the finite size
dependence of the energies $\En$, $\Ec$, $E_{\Phi}$ and $\ep_B$ we have
computed the $S^{(n)}$, $S^{(c)}$ and $G_{\Phi}$ propagators
at a fixed point in VXS phase on a sequence of lattices with
size $L \times 64$, where spatial extent $L$ was varied
in a range from 16 to 64.
The results for $\En$, $\Ec$, $E_{\Phi}$ and $\ep_B$ are summarized
in table 1.
\begin{table}
\begin{center}
\begin{tabular}{|c|c|c|c|c|c|} \hline
$L$
  & $E_{\Phi}$ & $\En$ & $\Ec$ & $\ep_B$ \\
\hline
16 &  0.126(3) & 0.275 & 0.368(13) & -0.033(16) \\ \hline
32 &  0.115(3) & 0.277 & 0.374(9)  & -0.018(12) \\ \hline
48 &  0.120(5) & 0.276 & 0.379(12) & -0.017(17) \\ \hline
64 &  0.119(5) & 0.277 & 0.389(11) & -0.007(16) \\ \hline
\end{tabular}
\end{center}
\caption{{\em The rest energies $E_{\Phi}$,
$\En$, $\Ec$ and $\ep_B$
as a function of the spatial extent $L$
of a lattice with volume $L \times 64$ obtained at the point
$(\kp,y,w)=(0.45,0.3,2.0)$. The error bars for $\En$ were omitted since
they are smaller than $0.001$.}}
\label{table:TAB1}
\end{table}
It can be seen also here that the binding energies are very small and
even on the smallest lattices almost
compatible with zero within the quoted error bars. Furthermore one
recognizes a systematic trend of $|\ep_B|$ to
decrease when the spatial extent of the lattice is enlarged.
These results strongly suggest that
the signal which we detected in the charged fermion propagators is
simply caused by a two
particle state of the neutral fermion
and the scalar particle. We now can understand also the fact why the
hopping expansion to lowest order
for the charged fermion propagator is very misleading. To
reproduce the inverse propagator of a two particle state an infinite
number of hopping terms would be needed.
Only on the basis of our numerical results
we can, of course, not completely exclude
the possibility that in principle a very small and non-vanishing
binding energy might be left over in the continuum limit.
We could also not find a good field theoretical argument
which would rule out the existence of a bound state
with zero binding energy. Both cases would
bring us back to scenario A which we drew up first in this
section.\\

Let us now proceed under the assumption that
the charged fermion does not exist as a particle in the spectrum.
In order to find an effective lagrangian for the model in the VXS phase
we first rewrite the lattice lagrangian (\eq{LN}) in the following way
\bea
{\cal L}_F &=& \half \sum_{\mu=1}^2 \left[
(\Pnb_{L,x}  \gm_{\mu} U^{\prime}_{\mu x} \Pn_{L,x+\hmu} -
\Pnb_{L,x+\hmu}  \gm_{\mu} {U^{\prime}_{\mu x}}^* \Pn_{L,x}) +
(\Pnb_{R,x} \gm_{\mu} \Pn_{R, x+\hmu} -
\Pnb_{R,x+\hmu} \gm_{\mu} \Pn_{R, x}) \right] \nonumber \\
& & +y \Pnb \Pn - \frac{w}{2} \Pnb \sum_{\mu=1}^2 \Pm \Pbm \Pn \;,\lb{LN1}
\eea
where we introduced the effective gauge field
\be
U_{\mu x}^{\prime} \equiv \Phi^*_x U_{\mu x} \Phi_{x+\hmu} \;. \lb{UPR}
\ee
Although the interaction in the first term of eq.~(\eq{LN1})
appears to be too weak for a formation of a $\Phi$-$\Pn$ bound
state, the form (\eq{LN1}) leaves  still the possibility of a chiral
coupling between the neutral fermion and the effective gauge field
$U_{\mu x}^{\prime}$ since this
field has dimension one according to a naive power counting analysis.
This outcome would be very interesting since one would have found at least one
example for a lattice regularized theory
where fermions exhibit a chiral coupling to an ``external'' gauge field.
In contrast in four dimensional models
the naive power counting analysis suggests
that the coupling to the effective gauge field $U_{\mu x}^{\prime}$
vanishes like $a^2$ in the continuum limit.\\

The U(1) pure gauge model (\eq{LGAUGE})
is confining for all values of the gauge coupling $g$.
The confinement phase is expected to be present also in the two dimensional
U(1) Higgs model at small values of $\kp$ and to turn
in the limit $g \ra 0$ into the vortex phase. For $g>0$ the scalar particles
get confined into massive bosonic particles. The effective gauge field
in eq.~(\eq{UPR}) can be written in the form
\be
U_{\mu x}^{\prime} =z^2 + H_{\mu x} + i W_{\mu x}  \lb{UEFF}
\ee
where $z^2$ is given in eq.~(\eq{LINK}) and $H_{\mu x}$ and $W_{\mu x}$
are interpolating fields for bosonic bound states in the
confinement phase with quantum numbers
$J^{PC}=0^{++}$ and $1^{--}$ in lowest spin state \cc{Evertz}.
The field $H_{\mu x}$ couples
primarily to the Higgs-like scalar particle according to
\be
H_{\mu x} \ra m_H \; H_x     \lb{HIGGS}
\ee
where $m_H$ is some mass scale which has to be introduced since the
scalar field $H_x$ is dimensionless
($H_{\mu x}$ is not a vector under lattice rotations,
which forbids a
relation of the form $H_{\mu x} \ra \pl_{\mu} H(x)$).
The field $W_{\mu x}$ couples to the vector boson. \\

After inserting (\eq{UEFF}) into (\eq{LN1}) and a trivial
rescaling of the fields $\Pn_L$ and $\Pnb_L$
we obtain for ${\cal L}_F$ the form
\bea
{\cal L}_F &=& \half \sum_{\mu=1}^2 \left[
               \Pnb_x  \gm_{\mu}  \Pn_{x+\hmu} -
               \Pnb_{x+\hmu}  \gm_{\mu}  \Pn_{x} \right]
    + \frac{y}{z} \Pnb \Pn - \frac{w}{2z} \Pnb \sum_{\mu=1}^2 \Pm \Pbm \Pn
      \lb{LN2A} \\
& & + \frac{m_H}{z^2} \half \sum_{\mu=1}^2 H_x \left[
      \Pnb_{L,x}  \gm_{\mu} \Pn_{L,x+\hmu} -
      \Pnb_{L,x+\hmu}  \gm_{\mu} \Pn_{L,x} \right] \lb{LN2B} \\
& & + \frac{1}{z^2} \half \sum_{\mu=1}^2 i W_{\mu x} \left[
      \Pnb_{L,x}  \gm_{\mu} \Pn_{L,x+\hmu} +
      \Pnb_{L,x+\hmu}  \gm_{\mu} \Pn_{L,x} \right] \lb{LN2C}
\eea
The term (\eq{LN2A}) describes a free neutral fermion with mass $\mn =y/z$.
The expression (\eq{LN2B}) suggests that the coupling of the neutral fermion
to the Higgs-like bound state vanishes like $a$. However,
the neutral fermion couples chirally in (\eq{LN2C}) to the
vector boson field $W_{\mu x}$ if its
dimension is one, as suggested by the naive dimensional analysis. \\

In order to find out whether the fields $W_{\mu x}$ and $H_{\mu x}$
are indeed dimension one operators we have computed the
scale dependence of the corresponding wave-function renormalization constants
$Z_H$ and $Z_W$. From the naive dimensional analysis
these wave-function renormalization constants $Z_H$ and $Z_W$
are expected to vanish like $a^2$. To see whether this expectation is
correct we have computed the momentum space propagators
\be
G_{H}(p_2)=\left\lag \frac{1}{V} \sum_{x,y}
H_{1 x} H_{1 y} e^{i p_2 (x_2-y_2)} \right\rag \;,\;\;
G_{W}(p_2)=\left\lag \frac{1}{V} \sum_{x,y}
W_{1 x} W_{1 y} e^{i p_2 (x_2-y_2)} \right\rag
\ee
in the confinement phase of the U(1) gauge-Higgs model
for several values of $g$ and $\kp$ and fitted the results
for sufficiently small $p_2$ to the free boson propagator ansatz
given in eq.~(\eq{FB}), which for $G_W$ is considered as a special
case ($p_1=0$) of a free massive vector boson propagator,
$(\delta_{\mu \nu} + p_{\mu}
p_{\nu}/m^2)(m^2+p^2)^{-1}$. In fig.~6 we have displayed
the resulting wave-function
renormalization constant $Z_W$ (squares)
and $Z_H$ (circles) respectively as a function of
$m^2=a^2 m_{W,phys}^2$ and $m^2=a^2 m_{H,phys}^2$
for the fixed ratio $m_H/m_W=1.14$. In both cases
the wave-function renormalization constants obey nicely
a linear dependence, supporting our expectation from the dimensional
analysis.            \\
%
%
\begin{figure}[tb]
\vspace{12cm}
\caption{ \noindent {\em
The wave-function renormalization constants $Z_W$ (squares)
and $Z_H$ (circles) are plotted respectively as a function of $m^2=m_W^2$
and $m=m_H^2$ for the fixed ratio $m_H/m_W=1.14$.
The two dotted lines through the origin
are drawn to guide
the eye.}}
\label{td6.ps}
\end{figure}

This result suggests that the neutral fermion exhibits
in two dimensions
indeed a non-vanishing chiral coupling to the massive vector bosons.
This coupling is universal in the quenched approximation
since the field $W_{\mu x}$ is proportional to a current which is
conserved in the two dimensional U(1) gauge-Higgs model.
The properties of this coupling in the full
model with dynamical fermions are, however, not yet clear to us. \\

The fermion couplings in the VXS phase may then be summarized
qualitatively by the following effective lagrangian
\be
{\cal L}_F^{eff} = \psbn \dsl \psn +\mn \psbn \psn
+ g_R \psbn_L \gm_{\mu} \psn_L  W_{\mu}^{(c)} \;, \lb{EFF2}
\ee
where $g_R=\sqrt{Z_{W}}/z^2$ and $W_{\mu}^{(c)}$ is the
vector field in the continuum with standard normalization. This
expression for the effective lagrangian gives a satisfactory description
of the fermion couplings at distances which are large in comparison with
the typical length scale of the vector boson bound state.
When lowering the value of the
gauge couplings $g$ the string tension becomes smaller and the bound
states extends over larger distances. At small distances the
scalar particles are then almost free and a more appropriate form of the action
is given then by
\be
{\cal L}_F^{eff} \ra  \psbn \dsl \psn +\mn \psbn \psn
+ \frac{1}{z^2} \psbn_L \gm_{\mu} \psn_L
(\phi^* \partial_{\mu}
\phi - \phi \partial_{\mu} \phi^*) \;.  \lb{EFF3}
\ee
This form is expected to describe the fermion couplings
in particular in the global limit, i.~e. $g=0$, where the
bosonic spectrum consists of unbound scalar particles.
The massive neutral fermion in eq.~(\eq{EFF3})
is coupled to a two particle current. This interaction seems,
however, to be quite weak since our numerical results for the neutral fermion
propagator are in good agreement with the analytic prediction from the
lowest order hopping expansion, which leads to a free fermion propagator.
Furthermore our numerical results for the binding energy $\ep_B$
suggest that the interaction in (\eq{EFF3}) is too weak as to give
rise to the formation of a $\Phi$-$\Pn$ bound state.
Also the expressions (\eq{EFF2}) and (\eq{EFF3}) are certainly different from
the target model in eq.~(\eq{CONT}) which we had originally in mind. \\
%
%
%
%
%
\section{Conclusion}
We started out our investigations from the lattice lagrangian given
in eq.~(\eq{S}) in the hope to obtain in the continuum limit
the target model in eq.~(\eq{CONT}). In the global limit of the model
($g\ra0$) the unwanted species doublers can be
removed completely from the spectrum within the strong coupling vortex
phase (VXS). The physics in this phase differs, however, substantially from
the target action which we had originally in mind. In the previous chapter
we have developed two different scenarios (A and B) for the effective theory
in the VXS phase which were summarized by the continuum
lagrangians (\eq{mirror}) and (\eq{EFF2}), (\eq{EFF3}).
Our numerical results are in favor of scenario B:
In this case the fermionic spectrum in VXS contains only
a neutral fermion which has zero charge with respect to the U(1)$_L$ group
and which in the global limit $g \ra 0$ exhibits a left-handed
coupling to a two particle current. This coupling, however,
appears to be weak since the neutral fermion propagator data are
in nice agreement with the results from the lowest
order hopping expansion which implies a free fermion behavior.
Furthermore this coupling seems to be too weak as to give rise to
the formation of a $\Phi$-$\Pn$ bound state. When the gauge coupling
is turned on, we argue that the neutral fermion couples
chirally to the massive vector boson state in the
confinement phase. If this scenario is correct, it would have been
the first time that a chirally coupled fermion has been
detected on the lattice. This result is also different from the previous
findings in the strong coupling symmetric phase (PMS)
of the fermion-Higgs models with Wilson-Yukawa
coupling in four dimensions where the coupling of the neutral fermion
to the bosonic bounds states vanishes presumably as a power of the
lattice spacing \cc{GoPe91b}. \\

\noindent {\bf Acknowledgements}\\
We would like to thank J.~Jers\'ak for reading the manuscript and for many
helpful comments. We have benefitted from discussions with
A.~Bochkarev, M.F.L.~Golterman, D.N.~Petcher
and J.~Vink. E.~F. and A.K.~D. thank
H.A.~Kastrup for his continuous support. The research was supported
by the ``Stichting voor Fundamenteel Onderzoek der Materie'' (FOM),
the ``Stichting Nationale Computer Faciliteiten (NCF)'' and by
the Deutsche Forschungsgemeinschaft. The numerical simulations were
performed on the CRAY Y-MP/832 at HLRZ J\"ulich and on the CRAY Y-MP4/464
at SARA, Amsterdam.
\end{document}